\documentclass[aps,superscriptaddress,eqsecnum,nofootinbib,preprintnumbers,twocolumn]{revtex4}
\pdfoutput=1

\usepackage{longtable}
\usepackage{bm}
\usepackage{relsize}
\usepackage{amsfonts}
\usepackage{amsmath}
\usepackage{amssymb,epsf}
\usepackage{latexsym}
\usepackage{graphicx,epsfig}
\usepackage{amssymb}
\usepackage{float}
\usepackage{subfigure}
\usepackage{epstopdf}
\usepackage[colorlinks=true,citecolor=blue,linkcolor=blue,urlcolor=black]{hyperref}
\usepackage{dcolumn}
\usepackage{psfrag}
\usepackage{wrapfig}
\usepackage{makeidx}
\usepackage{epsf}
\usepackage{color}
\usepackage{multirow}
\usepackage{mathtools}

\begin{document}

\title{Dynamical reconstruction of the $\Lambda$CDM model in hybrid metric-Palatini gravity}

\author{João Luís Rosa}
\email{joaoluis92@gmail.com}
\affiliation{Institute of Physics, University of Tartu, W. Ostwaldi 1, 50411 Tartu, Estonia}
\affiliation{Institute of Theoretical Physics and Astrophysics, University of Gda\'{n}sk, Jana Ba\.{z}y\'{n}skiego 8, 80-309 Gda\'{n}sk, Poland}

\date{\today}

\begin{abstract} 
In this work, we apply the formalism of dynamical systems to analyze the viability of the $\Lambda$CDM model in a generalized form of the hybrid metric-Palatini gravity theory written in terms of its dynamically equivalent scalar-tensor representation. Adopting a matter distribution composed of two relativistic fluids described by the equations of state of radiation and pressureless dust, one verifies that the cosmological phase space features the usual curvature-dominated, radiation-dominated, matter-dominated, and exponentially accelerated fixed points, even in the absence of a dark energy component. A numerical integration of the dynamical equations describing the system, subjected to initial conditions consistent with the cosmographic observations from the Planck satellite and weak-field solar system dynamics, shows that cosmological solutions with the same behavior as the $\Lambda$CDM model in General Relativity (GR) are attainable in this theory, with the deviations from GR being exponentially suppressed at early-times and the scalar-field potential effectively playing the role of dark energy at late times. 
\end{abstract}

\maketitle

\section{Introduction}\label{sec:intro}

According to the current cosmological measurements, the universe is undergoing a phase of accelerated expansion \cite{SupernovaCosmologyProject:1998vns,SupernovaSearchTeam:1998fmf,Planck:2018vyg}. In the framework of General Relativity (GR), this phenomenon is commonly attributed to an unknown matter component described by a relativistic fluid with a negative pressure known as dark energy \cite{Copeland:2006wr,Li:2011sd,Peebles:2002gy}. Despite the apparent success of dark energy models, a viable alternative to explain these observations lies on modifications of the gravitational theory itself \cite{Clifton:2011jh,Capozziello:2011et,Nojiri:2017ncd,Nojiri:2010wj}. One of the most popular modifications of GR is the $f\left(R\right)$ gravity \cite{Sotiriou:2008rp,DeFelice:2010aj}, which although allowing for accelerating cosmological behaviors without the need for dark energy sources, and successfully addressed the dynamics of gravitating systems in the presence of dark matter \cite{Boehmer:2007kx}, feature severe drawbacks in terms of weak-field dynamics \cite{Khoury:2003aq,Khoury:2003rn,Dolgov:2003px,Sotiriou:2006sf}, and other cosmological issues \cite{Amendola:2006kh,Amendola:2006we,Tsujikawa:2007tg}. To overcome these undesirable effects, a Palatini formulation of the $f\left(R\right)$ gravity was proposed \cite{Olmo:2011uz}, but raises other issues related to the physics of compact stars \cite{Kainulainen:2006wz} and cosmological perturbations \cite{Koivisto:2005yc,Koivisto:2006ie}.

A hybrid version of the $f\left(R\right)$ theory encompassing elements from both the metric and the Palatini formalism, i.e., a hybrid metric-Palatini gravity \cite{Capozziello:2015lza}, was shown to successfully unify late-time cosmic acceleration with solar system constraints \cite{Harko:2011nh,Capozziello:2013uya}, as well as several other issues in the fields of cosmology and astrophysics \cite{Capozziello:2012ny,Capozziello:2012qt,Koivisto:2013kwa,Capozziello:2012hr,Capozziello:2013yha,Carloni:2015bua,Edery:2019txq,Harko:2020ibn,Danila:2016lqx,Danila:2018xya,Bronnikov:2019ugl,Leanizbarrutia:2017xyd,Bekov:2020dww,Chen:2020evr}. The theory was further generalized to include an arbitrary dependence on on the Ricci scalars from the metric and Palatini formalisms \cite{Tamanini:2013ltp}, allowing to further extend the range of applicability of the hybrid metric-Palatini gravity \cite{Rosa:2018jwp,Rosa:2017jld,Rosa:2020uoi,Rosa:2019ejh,Borowiec:2020lfx,Rosa:2021yym,Rosa:2020uli,Bombacigno:2019did,Rosa:2021mln,Sa:2020qfd,Rosa:2021lhc,Rosa:2021ish,daSilva:2021dsq,Bronnikov:2021tie,Golsanamlou:2023wiz}. The main goal of this work is thus to extend on the current cosmological literature and to analyze the cosmological phase space of the theory through the use of the formalism of dynamical systems \cite{Bahamonde:2017ize}.

The dynamical system approach allows one to analyze how the different quantities describing a model evolve through time as a whole. The versatility of these methods led to a wide plethora of applications in the field of cosmology in the framework of modified theories of gravity \cite{Odintsov:2017tbc,Carloni:2015jla,Alho:2016gzi,Carloni:2007eu,Rosa:2023qun,Carloni:2017ucm,Carloni:2009jc,Carloni:2015lsa,Carloni:2018yoz,Carloni:2007br,Carloni:2013hna,Bonanno:2011yx,Goncalves:2023klv}. Indeed, the dynamical system formalism has already been used to analyze the cosmological phase space of hybrid metric-Palatini gravity \cite{Carloni:2015bua,Tamanini:2013ltp,Rosa:2019ejh}. However, these works feature several drawbacks, e.g. the lack of an explicit comparison with experimental results, and the necessity to impose a particular form of the theory \textit{a priori}, thus resulting in theory-dependent solutions. We thus aim to suppress this gap in the literature by providing a theory-independent cosmological model consistent simultaneously with the cosmological measurements of the Planck satellite and weak field solar system dynamics.

This manuscript is organized as follows. In Sec. \ref{sec:theory} we introduce the hybrid metric-Palatini gravity theory in both its geometrical and scalar-tensor representations, and we introduce the framework and assumptions for the geometry and matter distributions; in Sec. \ref{sec:dynsys} we introduce the dynamical system formalism, define a set of dimensionless dynamical variables and deduce the set of dynamical equations describing the theory. Furthermore, we analzye the structure of the phase space in terms of critical points and phase space trajectories, and we perform a full numerical integration of the dynamical system subjected to appropriate initial conditions in order to obtain a cosmological model compatible with the $\Lambda$CDM model\footnote{The acronym $\Lambda$CDM stands for the cosmological constant $\Lambda$ and Cold Dark Matter.}; and in Sec. \ref{sec:concl} we summarize our results and trace our conclusions.

\section{Theory and framework}\label{sec:theory}

\subsection{Hybrid metric-Palatini gravity}

The generalized form of the hybrid metric-Palatini theory if gravity is described by an action $S$ of the form
\begin{equation}\label{eq:gaction}
S=\frac{1}{2\kappa^2}\int_\Omega\sqrt{-g}f\left(R,\mathcal R\right)d^4x+S_m,
\end{equation}
where $\kappa^2\equiv 8\pi G/c^4$, where $G$ is the gravitational constant and $c$ is the speed of light, $\Omega$ is the four-dimensional spacetime manifold, $g$ is the determinant of the metric $g_{\mu\nu}$ written in terms of a coordinate system $x^\mu$, $f\left(R,\mathcal R\right)$ is an arbitrary well-behaved function of the two scalars, $R\equiv g^{\mu\nu} R_{\mu\nu}$ is the Ricci scalar of the metric $g_{\mu\nu}$ built in terms of the Levi-Civita connection $\Gamma^\alpha_{\mu\nu}$, with $R_{\mu\nu}$ the corresponding Ricci tensor, and $\mathcal R\equiv g^{\mu\nu}\mathcal R_{\mu\nu}$ is the Palatini Ricci scalar, with $\mathcal R_{\mu\nu}$ the Palatini Ricci tensor, which is defined in terms of an independent metric-incompatible connection $\hat\Gamma^\alpha_{\mu\nu}$ as
\begin{equation}\label{eq:palatiniRab}
    \mathcal R_{\mu\nu}=\partial_\alpha\hat\Gamma^\alpha_{\mu\nu}-\partial_\nu\hat\Gamma^\alpha_{\mu\alpha}+\hat\Gamma^\alpha_{\alpha\beta}\hat \Gamma^\beta_{\mu\nu}-\hat\Gamma^\alpha_{\mu\beta}\hat\Gamma^\beta_{\alpha\nu},
\end{equation}
where $\partial_\mu$ denotes a partial derivative with respect to the coordinate $x^\mu$. The metric incompatibility of the independent connection implies that $\hat\nabla_\alpha g_{\mu\nu}\neq 0$, where $\hat\nabla_\mu$ denotes the covariant derivatives written in terms of the connection $\hat\Gamma^\alpha_{\mu\nu}$. Furthermore, although it is not strictly necessary, throughout this manuscript we assume that the independent connection is torsionless, i.e., $\hat\Gamma^\alpha_{\mu\nu}=\hat\Gamma^\alpha_{\nu\mu}$. Finally, $S_m\equiv \int_\Omega\sqrt{-g}\mathcal L_m d^4x$ is the matter action, with $\mathcal L_m$ the matter Lagrangian density, which is assumed to be minimally coupled to the gravitational sector. In what follows, we adopt a system of geometrized units in such a way that $G=c=1$, which implies consequently that $\kappa^2=8\pi$.  

The action in Eq. \eqref{eq:gaction} depends explicitly on two independent quantities, namely the metric $g_{\mu\nu}$ and the connection $\hat\Gamma^\alpha_{\mu\nu}$. A variation of the action with respect to the metric yields the modified field equations of the theory, given by
\begin{eqnarray}\label{eq:gfield}
f_R R_{\mu\nu}+f_\mathcal{R}\mathcal R_{\mu\nu}-\frac{1}{2}g_{\mu\nu}f\left(R,\mathcal R\right)-\nonumber \\
-\left(\nabla_\mu\nabla_\nu -g_{\mu\nu}\Box \right)f_R=8\pi T_{\mu\nu},
\end{eqnarray}
where we have introduced the notation $f_R\equiv \partial f/\partial R$ and $f_\mathcal R\equiv \partial f/\partial\mathcal R$, $\nabla_\mu$ denotes the covariant derivatives written in terms of the Levi-Civita connection $\Gamma^\alpha_{\mu\nu}$, $\Box=g^{\mu\nu}\nabla_\mu\nabla_\nu$ is the d'Alembert operator, and $T_{\mu\nu}$ is the stress-energy tensor defined in terms of a variation of the matter Lagrangian as
\begin{equation}\label{eq:defTab}
    T_{\mu\nu}=-\frac{2}{\sqrt{-g}}\frac{\delta\left(\sqrt{-g}\mathcal L_m\right)}{\delta g^{\mu\nu}}.
\end{equation}
Taking instead a variation with respect to the connection $\hat\Gamma^\alpha_{\mu\nu}$, one obtains the equation of motion
\begin{equation}\label{eq:eomhatgamma}
\hat\nabla_\alpha\left(\sqrt{-g}f_\mathcal R g_{\mu\nu}\right)=0.
\end{equation}
Given that the scalar $\sqrt{-g}$ is a scalar density of weight $1$, one obtains that $\hat\nabla_\mu \sqrt{-g}=0$ and thus Eq. \eqref{eq:eomhatgamma} can be further simplified into $\hat\nabla_\alpha\left(f_\mathcal R g_{\mu\nu}\right)=0$. The latter result implies that there exists a metric tensor $\hat g_{\mu\nu}\equiv f_\mathcal R g_{\mu\nu}$ conformally related to the metric $g_{\mu\nu}$ such that the independent connection $\hat\Gamma^\alpha_{\mu\nu}$ is the Levi-civita connection of the metric $\hat g_{\mu\nu}$, i.e., one can write
\begin{equation}
\hat\Gamma^\alpha_{\mu\nu}=\frac{1}{2}\hat g^{\alpha\beta}\left(\partial_\mu \hat g_{\beta\nu}+\partial_\nu \hat g_{\mu\beta}-\partial_\beta \hat g_{\mu\nu}\right).
\end{equation}
The conformal relation between the metrics $\hat g_{\mu\nu}$ and $g_{\mu\nu}$ implies that there is a conformal relation between the Ricci tensor $R_{\mu\nu}$ and the Palatini Ricci tensor $\mathcal R_{\mu\nu}$, which is given by
\begin{equation}\label{eq:Rabrel}
    \mathcal R_{\mu\nu}=R_{\mu\nu}-\frac{1}{f_\mathcal R}\left(\nabla_\mu\nabla_\nu+\frac{1}{2}g_{\mu\nu}\Box\right)f_\mathcal R+\frac{3}{2 f^2_\mathcal R}\partial_\mu f_\mathcal R\partial_\nu f_\mathcal R.
\end{equation}
Note that Eqs. \eqref{eq:eomhatgamma} and \eqref{eq:Rabrel} are equivalent, and thus we recur to the latter due to its more convenient structure for the analysis that follows.

\subsection{Scalar-tensor representation}

Under certain circumstances, it is convenient to rewrite the action and equations of motion of the theory in terms of a dynamically equivalent scalar-tensor representation with two scalar fields. Such a transformation reduces the order of the field equations from fourth to second degree, at the cost of introducing two additional equations of motion for the scalar fields. To perform this transformation, consider the gravitational part of the action in Eq. \eqref{eq:gaction} conveniently rewritten in terms of two auxiliary fields $\alpha$ and $\beta$ as
\begin{equation}\label{eq:auxaction}
    S=\frac{1}{2\kappa^2}\int_\Omega\sqrt{-g}\left[f\left(\alpha,\beta\right)+f_\alpha\left(R-\alpha\right)+f_\beta\left(\mathcal R-\beta\right)\right]d^4x,
\end{equation}
where we have introduced the notation $f_\alpha\equiv\partial f/\partial\alpha$ and $f_\beta \equiv\partial f/\partial\beta$. Equation \eqref{eq:auxaction} depends explicitly on three independent quantities, namely the metric $g_{\mu\nu}$ and the auxiliary fields $\alpha$ and $\beta$. Taking a variation of Eq. \eqref{eq:auxaction} with respect to the auxiliary fields $\alpha$ and $\beta$, one verifies that the resultant equations of motion can be written in a matrix form as
\begin{equation}\label{eq:matrixeq}
    \mathcal M \textbf{x}\equiv\begin{pmatrix} f_{\alpha\alpha} & f_{\alpha\beta} \\ f_{f_\beta\alpha} & f_{\beta\beta}\end{pmatrix} \begin{pmatrix} R-\alpha \\ \mathcal R-\beta \end{pmatrix} = 0.
\end{equation}
The solution of Eq. \eqref{eq:matrixeq} is unique if and only if the determinant of the matrix $\mathcal M$ is non-zero, i.e., $f_{\alpha\alpha}f_{\beta\beta}-f_{\alpha\beta}^2\neq 0$, where we have assumed that the function $f\left(\alpha,\beta\right)$ satisfied the Schwarz theorem. In such a case, the unique solution is $\alpha=R$ and $\beta=\mathcal R$, and consequently Eqs. \eqref{eq:auxaction} and \eqref{eq:gaction} are equivalent. The non-vanishing of the determinant of $\mathcal M$ is essential to guarantee that the scalar-tensor representation of the theory is well-defined. Introducing the following definitions of the scalar fields and interaction potential:
\begin{equation}\label{eq:defscalar}
    \varphi=f_\alpha,\qquad \psi=-f_\beta, 
\end{equation}
\begin{equation}\label{eq:defpotential}
    V\left(\varphi,\psi\right)=-f\left(\alpha,\beta\right)+\varphi\alpha-\psi\beta,
\end{equation}
the action in Eq. \eqref{eq:auxaction} can be rewritten in terms of $\varphi$, $\psi$ and $V$ as
\begin{equation}\label{eq:auxaction2}
    S=\frac{1}{2\kappa^2}\int_\Omega\sqrt{-g}\left[\varphi R-\psi\mathcal R-V\left(\varphi,\psi\right)\right]d^4x.
\end{equation}
Finally, taking into consideration that $R_{\mu\nu}$ and $\mathcal R_{\mu\nu}$ are conformally related via Eq. \eqref{eq:Rabrel}, contracting with $g^{\mu\nu}$ and introducing the definition of $\psi$ from Eq. \eqref{eq:defscalar}, one obtains a relation between the Ricci scalar $R$ and the Palatini Ricci scalar $\mathcal R$ of the form
\begin{equation}\label{eq:Rrel}
    \mathcal R=R-\frac{3}{\psi}\Box\psi+\frac{3}{2\psi^2}\partial^\mu\psi\partial_\mu\psi,
\end{equation}
which can be introduced into Eq. \eqref{eq:auxaction2} to effectively remove the scalar $\mathcal R$ from the action. The result is the action describing the scalar-tensor representation of the theory, which takes the form
\begin{eqnarray}\label{eq:staction}
    S&=&\frac{1}{2\kappa^2}\int_\Omega\sqrt{-g}\left[\left(\varphi-\psi\right)R-\right.\nonumber \\
    &-&\left.\frac{3}{2\psi}\partial^\mu\psi\partial_\mu\psi-V\left(\varphi,\psi\right)\right]d^4x+S_m.
\end{eqnarray}
We note that the term proportional to $\Box\psi$ in Eq. \eqref{eq:Rrel} is absent from Eq. \eqref{eq:staction} due to the fact that it can be written as a boundary term via the Stokes theorem, and thus it does not contribute to the equations of motion. 

Equation \eqref{eq:staction} depends explicitly on three independent quantities, namely the metric $g_{\mu\nu}$ and the scalar fields $\varphi$ and $\psi$. A variation with respect to $g_{\mu\nu}$ yields the modified field equations in the scalar-tensor representation, which take the form
\begin{eqnarray}
    \left(\varphi-\psi\right)G_{\mu\nu}&=&\kappa^2T_{\mu\nu}+\frac{3}{2\psi}\partial_\mu\psi\partial_\nu\psi-\frac{3}{4\psi}g_{\mu\nu}\partial^\alpha\psi\partial_\alpha\psi-\nonumber \\
    &-&\frac{1}{2}g_{\mu\nu}V+\left(\nabla_\mu \nabla_\nu-g_{\mu\nu}\Box\right)\left(\varphi-\psi\right).\label{eq:stfield}
\end{eqnarray}
Finally, taking a variation of Eq. \eqref{eq:staction} with respect to the scalar fields $\varphi$ and $\psi$ yields the equations of motion 
\begin{equation}\label{eq:eomphi}
    V_\varphi = R,
\end{equation}
\begin{equation}\label{eq:eompsi}
    V_\psi = -R+\frac{3}{\psi}\Box\psi-\frac{3}{2\psi^2}\partial^\mu\psi\partial_\mu\psi,
\end{equation}
where we have introduced the notation $V_\varphi\equiv \partial V/\partial\varphi$ and $V_\psi \equiv \partial V/\partial_\psi$. In what follows, we work in the scalar-tensor representation of the theory, i.e., the equations of interest are Eqs. \eqref{eq:stfield} to \eqref{eq:eompsi}.

\subsection{geometry and matter distribution}

We are interested in analyzing the hybrid metric-Palatini gravity theory in a cosmological context. For this purpose, we assume that the spacetime is well-described by a homogeneous and isotropic Friedmann-Lemaître-Robertson-Walker (FLRW) metric, whose line-element is given in the usual system of spherical coordinates $x^\mu=\left(t,r,\theta,\phi\right)$ as
\begin{equation}\label{eq:metric}
    ds^2=-dt^2+a^2\left(t\right)\left(\frac{dr^2}{1-k r^2}+r^2 d\Omega^2\right),
\end{equation}
where $a\left(t\right)$ is the scale factor which is assumed to be a function only of the time coordinate in order to preserve the homogeneity of the spacetime, $k$ is the sectional curvature of the spacetime which takes the values $+1$, $-1$, and $0$ for spherical, hyperbolic, and flat geometries, respectively, and $d\Omega^2=d\theta^2+\sin^2\theta d\phi^2$ is the line-element on the two-sphere. In what follows, we introduce the Hubble parameter defined as
\begin{equation}\label{eq:defH}
    H=\frac{\dot a}{a},
\end{equation}
where a dot $\dot\ $ denotes a derivative with respect to the time coordinate $t$.

Regarding the matter sector, we assume that the matter distribution is well-described by a superposition of isotropic perfect fluids, i.e., the stress-energy tensor can be written in the diagonal form
\begin{equation}\label{eq:tab}
    g^{\mu\nu}T_{\mu\nu}=\text{diag}\left(-\rho,p,p,p\right),
\end{equation}
where $\rho\equiv\rho\left(t\right)$ denotes the total energy density and $p\equiv p\left(t\right)$ denotes the total isotropic pressure, both assumed to be functions of time only to preserve the homogeneity of the spacetime. The total energy density and the total pressure are the sum of the energy densities and isotropic pressures of the individual components of the matter distribution, i.e., $\rho=\sum_i \rho_i$ and $p=\sum_i p_i$, with the quantities $\rho_i$ and $p_i$ being related by equations of state of the form $p_i =w_i \rho_i$. For the purpose of this work and to preserve a consistency with experimental observations, we consider a superposition of two fluids $i=\{r,m\}$, the first corresponding to radiation and satisfying an equation of state of the form $p_r = \frac{1}{3}\rho_r$, and the second corresponding to pressureless dust and satisfying the equation of state $p_m=0$. 

Taking a covariant derivative of Eq. \eqref{eq:stfield} leads to the equation for the matter conservation of the form
\begin{equation}
    \nabla_\mu T^{\mu\nu}\equiv \sum_i\nabla_\mu T^{\mu\nu}_i=0,
\end{equation}
where we have defined $T^{\mu\nu}_i$ as the stress-energy tensor corresponding to the matter component $i$. Although the conservation equation requires only a conservation of the total stress-energy tensor defined in Eq. \eqref{eq:tab}, in what follows we assume that the two components are independently conserved, i.e., that there are no transformations of matter between radiation and dust. Following these assumptions, and taking into consideration the metric given in Eq. \eqref{eq:metric}, the conservation equations for radiation and dust take the forms
\begin{equation}\label{eq:consr}
    \dot \rho_r +4H\rho_r = 0,
\end{equation}
\begin{equation}\label{eq:consm}
    \dot \rho_m +3H\rho_m = 0.
\end{equation}

Following the assumptions outlined above for the spacetime geometry and distribution of matter, the two independent components of the modified field equations in Eq. \eqref{eq:stfield}, as well as the equations of motion for the scalar fields given in Eqs. \eqref{eq:eomphi} and \eqref{eq:eompsi}, take the forms
\begin{eqnarray}\label{eq:stfield1}
    &&\left(\varphi-\psi\right)\left(H^2+\frac{k}{a^2}\right)=\frac{8\pi}{3}\left(\rho_r+\rho_m\right)-\nonumber \\
    &&-H\left(\dot\varphi-\dot\psi\right)+\frac{1}{4}\frac{\dot\psi^2}{\psi}+\frac{V}{6},
\end{eqnarray}
\begin{eqnarray}\label{eq:stfield2}
    &&\left(\varphi-\psi\right)\left(2\dot H+3H^2+\frac{k}{a^2}\right)=-\frac{8\pi}{3}\rho_r-\nonumber \\
    &&-2H\left(\dot\varphi-\dot\psi\right)- \left(\ddot \varphi-\ddot\psi\right)-\frac{3}{4}\frac{\dot\psi}{\psi}+\frac{V}{2},
\end{eqnarray}
\begin{equation}\label{eq:steomphi}
    V_\varphi=6\left(\dot H+2H^2+\frac{k}{a^2}\right),
\end{equation}
\begin{equation}\label{eq:steompsi}
    V_\psi = -6\left(\dot H+2H^2+\frac{k}{a^2}\right)+\frac{3}{\psi}\left(\frac{\dot\psi^2}{2\psi}-3H\dot\psi-\ddot\psi\right).
\end{equation}

The system of Eqs.\eqref{eq:consr} to \eqref{eq:steompsi} for a system of six coupled ordinary differential equations of which only five are linearly independent. This can be proven by taking a time derivative of Eq. \eqref{eq:stfield1}, using Eq. \eqref{eq:stfield2} to replace $\dot H$, using Eqs. \eqref{eq:consr} and \eqref{eq:consr} to replace the terms $\dot \rho_r$ and $\dot \rho_m$ respectively, using Eqs. \eqref{eq:steomphi} and \eqref{eq:steompsi} to replace the terms $V_\varphi$ and $V_\psi$, from which one recovers Eq. \eqref{eq:stfield1}. Thus, one equation can be discarded from the system. Due to its more complicated structure, we opt for discarding Eq. \eqref{eq:stfield2}. Furthermore, via the use of a chain rule for the potential $V$, i.e., $\dot V = V_\varphi \dot\varphi + V_\psi\dot\psi$, and the quantities $V_\varphi$ and $V_\psi$ given in Eqs. \eqref{eq:steomphi} and \eqref{eq:steompsi}, respectively, one verifies that
\begin{eqnarray}\label{eq:steomV}
    \dot V = 6\left(\dot\varphi-\dot\psi\right)\left(\dot H+2H^2+\frac{k}{a^2}\right)+\nonumber \\
    +3\frac{\dot\psi}{\psi}\left(\frac{\dot\psi^2}{2\psi}-3H\dot\psi-\ddot\psi\right)
\end{eqnarray}
Since the partial derivatives of $V$ do not appear in any other equation in the system besides the equations of motion for the scalar fields, Eq. \eqref{eq:steomV} can effectively replace Eqs. \eqref{eq:steomphi} and \eqref{eq:steompsi}, thus reducing the number of independent equations in the system. One is thus left with a system of four independent equations, namely Eqs. \eqref{eq:consr}, \eqref{eq:consm}, \eqref{eq:stfield1}, and \eqref{eq:steomV}.

\section{Dynamical systems}\label{sec:dynsys}

\subsection{Dynamical variables and equations}

To analyze the system of equations obtained previously through the formalism of dynamical systems, one must define a set of dimensionless dynamical variables that describe the quantities of interest in the system. We thus introduce the following definitions for the dynamical variables:
\begin{equation}
    K=\frac{k}{a^2H^2}, \qquad \Omega_r=\frac{8\pi\rho_r}{3H^2}, \qquad \Omega_m=\frac{8\pi\rho_m}{3H^2},\nonumber 
\end{equation}
\begin{equation}
    \Phi=\varphi,\qquad \Psi=\psi, \qquad U=\frac{V}{6H^2}.
\end{equation}
Note that the quantities $K$, $\Omega_r$ and $\Omega_m$ are the usual curvature, radiation density, and matter density parameters commonly used in GR, whereas the additional quantities $\Phi$, $\Psi$ and $U$ describe the additional contributions of hybrid metric-Palatini gravity. 

It is also useful in what follows to introduce the definition of the deceleration parameter $Q$ 
\begin{equation}\label{eq:defQ}
    Q=-\frac{\ddot a}{aH^2}.
\end{equation}
This parameter is directly associated with the cosmological behavior of the scale factor. Indeed, for a particular point in the cosmological phase space for which the cosmological solution presents a constant value of $Q=Q_0$, the behavior of the scale factor can be extracted by a direct integration of Eq. \eqref{eq:defQ}, and takes the form
\begin{equation}
    a\left(t\right)=a_0\left[1+H_0\left(1+Q_0\right)\left(t-t_0\right)\right]^{\frac{1}{1+Q_0}},
\end{equation}
where we have introduced the initial conditions $a\left(t=t_0\right)=a_0$ and $H\left(t=t_0\right)=H_0$, where $t_0$ is the present age of the universe. Note that in the limit $Q_0\to -1$, the cosmological behavior follows an exponential expansion $a\left(t\right)=a_0\exp\left[H_0\left(t-t_0\right)\right]$ by virtue of the limit $\lim_{n\to\infty}\left(1+\frac{x}{n}\right)^n=\exp\left(x\right)$. This analysis also allows one to extract the instantaneous behavior of the scale factor at some point of the phase space for which the value of $Q$ is known. 

The analysis of the evolution of the dynamical system requires the use of a dimensionless time parameter. For the purpose of this work, we consider the number of e-folds $N\equiv \log\left(\frac{a}{a_0}\right)$, where $a_0$ is the scale factor of the present universe, as a dimensionless time parameter. The dynamical equations for the variables defined above are obtained through a differentiation with respect to $N$, which is related to a differentiation with respect to $t$ for an arbitrary quantity $X$ through
\begin{equation}
    X'\equiv \frac{dX}{dN}=\frac{1}{H}\frac{dX}{dt}=\frac{\dot X}{H}.
\end{equation}
Following the definitions outlined above, the system of equations in Eqs. \eqref{eq:consr}, \eqref{eq:consm}, \eqref{eq:stfield1}, and \eqref{eq:steomV} provide dynamical equations for the quantities $\Omega_r$, $\Omega_m$, $\Phi$, $\Psi$, and $U$. A dynamical equation for $K$ can be obtained by computing the derivative $K'$. The resultant set of dynamical equations takes the form
\begin{equation}\label{eq:dynfield}
    \left(\Phi-\Psi\right)\left(1+K\right)+\Phi'-\Psi'-\frac{\psi'^2}{4\Psi}=\Omega_r+\Omega_m+U,
\end{equation}
\begin{equation}\label{eq:dynK}
    K'=2QK,
\end{equation}
\begin{equation}\label{eq:dynR}
    \Omega_r'=2\Omega_r\left(Q-1\right),
\end{equation}
\begin{equation}\label{eq:dynM}
    \Omega_m'=\Omega_m\left(2Q-1\right),
\end{equation}
\begin{eqnarray}\label{eq:dynU}
    U'=2U\left(Q+1\right)+\left(\Phi'-\Psi'\right)\left(1-Q+K\right)-\nonumber \\
    -\frac{\Psi'^2}{4\Psi}\left(4-2Q-\frac{\Psi'}{\Psi}+2\frac{\Psi''}{\Psi'}\right).
\end{eqnarray}
The set of Eqs. \eqref{eq:dynK} to \eqref{eq:dynM} allow us to identify the presence of three invariant submanifolds, described by the constraints $K=0$, $\Omega_r=0$, and $\Omega_m=0$, respectively. Thus, any global property of the phase space must lie in the intersection of these three submanifolds. 

\subsection{Phase space and critical points}

Equations \eqref{eq:dynfield} to \eqref{eq:dynU} define the dynamical system that characterizes the scalar-tensor representation of the hybrid metric-Palatini gravity. The critical points in the cosmological phase space can be obtained by imposing the conditions $K'=\Omega'_r=\Omega'_m=U'=\Phi'=\Psi'=\Psi''=0$ on the dynamical system and solving the resultant constraints for the dynamical variables. A summary of the critical points in the phase space is given in Table \ref{tab:points}. The critical points are identified by letters corresponding to a given cosmological behavior, namely $\mathcal A$ for radiation dominated points, $\mathcal B$ for matter dominated points, $\mathcal C$ for exponentially accelerated points, $\mathcal D$ for curvature dominated (linearly expanding) points, and $\mathcal E$ for points with arbitrary behaviors. 

\begin{table}[]
    \centering
    \begin{tabular}{c|c c c c c c c}
         & $K$ & $\Omega_r$ & $\Omega_m$ & $U$ & $\Phi$ & $\Psi$ & $Q$ \\ \hline 
        $\mathcal A_1$ & 0 & $\Phi-\Psi$ & 0 & 0 & ind. & ind. & 1 \\
        $\mathcal A_2$ & 0 & $\Phi-U$ & 0 & ind. & ind. & 0 & 1 \\
        $\mathcal B_1$ & 0 & 0 & $\Phi-\Psi$ & 0 & ind. & ind. & $\frac{1}{2}$ \\
        $\mathcal B_2$ & 0 & 0 & $\Phi-U$ & ind. & ind. & 0 & $\frac{1}{2}$ \\
        $\mathcal C$ & 0 & 0 & 0 & $\Phi-\Psi$ & ind. & ind. & -1 \\
        $\mathcal D_1$ & -1 & 0 & 0 & 0 & ind. & ind. & 0 \\
        $\mathcal D_2$ & ind. & 0 & 0 & $\Phi\left(1+K\right)$ & ind. & 0 & 0 \\
        $\mathcal D_3$ & ind. & 0 & 0 & 0 & ind. & $\Phi$ & 0 \\
        $\mathcal E_1$ & 0 & 0 & 0 & $\Phi$ & ind. & 0 & ind. \\
        $\mathcal E_1$ & 0 & 0 & 0 & 0 & ind. & $\Phi$ & ind. \\
    \end{tabular}
    \caption{Critical points of the phase space described by the dynamical system of Eqs. \eqref{eq:dynfield} to \eqref{eq:dynU}. The tag "ind." denotes an independent quantity. }
    \label{tab:points}
\end{table}

The phase space of hybrid metric-Palatini gravity features several additional fixed points with respect to standard GR, some of which converging into the usual GR critical point in the appropriate GR limit. Indeed, the GR limit is obtained through $\Phi-\Psi=1$ and for both $\Phi$ and $\Psi$ constant, which consequently implies that the density parameters are constrained through the relation $1+K=\Omega_r+\Omega_m+U$ from Eq. \eqref{eq:dynfield}. In this limit, one thus observes that both points $\mathcal A_1$ and $\mathcal A_2$ converge to a situation with $\Omega_r=1$, that both points $\mathcal B_1$ and $\mathcal B_2$ converge to a situation with $\Omega_m=1$, and that both points $\mathcal D_1$ and $\mathcal D_2$ converge to a situation with $K=-1$ and $U=0$, as expected in the GR limit. Nevertheless, there are additional critical points that do not have a correspondence to GR. Indeed, points $\mathcal D_3$ and $\mathcal E_2$ can not be continuously linked to GR since they require $\Phi=\Psi$, which violates the condition $\Phi-\Psi=1$. Finally, point $\mathcal E_1$ reduces to point $\mathcal C$ in the GR limit.

Due to the large dimensionality of the dynamical system, the analysis of the stability of the critical points obtained and production of adequate streamplots of the phase space require the use of projections into the invariant submanifolds. Furthermore, to allow for a comparison of the results with the GR limit, we chose to perform this analysis within the projection $\Phi=\Psi+1$, which contains the GR limit as a particular case $\Psi=\Psi_0$ for any constant $\Psi_0$. Under this projection, Eq. \eqref{eq:dynfield} reduces to
\begin{equation}\label{eq:simpY}
    \Psi'^2=-4\Psi\left(1+K-\Omega_r-\Omega_m-U\right)\equiv 4\Psi X,
\end{equation}
where we have defined an auxiliary quantity $X=1+K-\Omega_r-\Omega_m-U$. This equation controls the deviation from GR. Indeed, if $\Psi'=0$, which happens for either $\Psi=0$ or $X=0$, one recovers the GR limit. Furthermore, $\Psi=0$ is an invariant submanifold, and thus any initial condition starting from that submanifold is constrained to evolve within it. To analyze the effect of deviations from GR, in what follows we adopt the projections $X=X_0=\{-0.1; 0; 0.1\}$. Following the definition of $X$ in Eq. \eqref{eq:simpY}, for each value of $X_0$ one obtains a constraint that allows one to eliminate one dynamical variable from the system. In this case, we chose to eliminate the variable $U$ in order to preserve the usual GR density and curvature parameters $\Omega_r$, $\Omega_m$, and $K$.

\begin{figure*}
    \centering
    \includegraphics[scale=0.53]{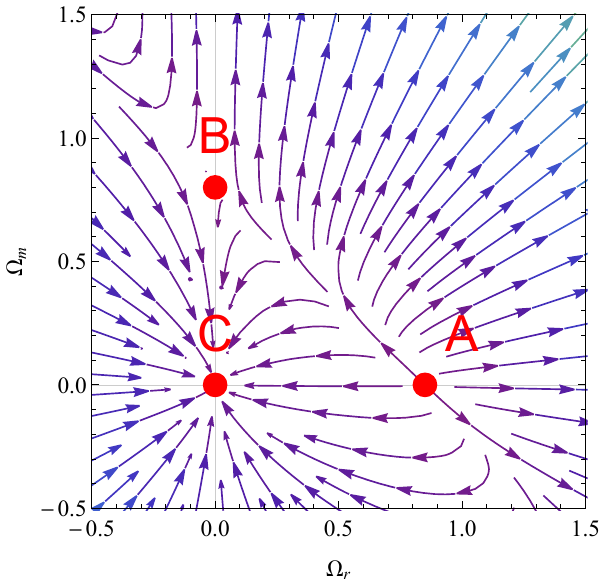} \quad
    \includegraphics[scale=0.53]{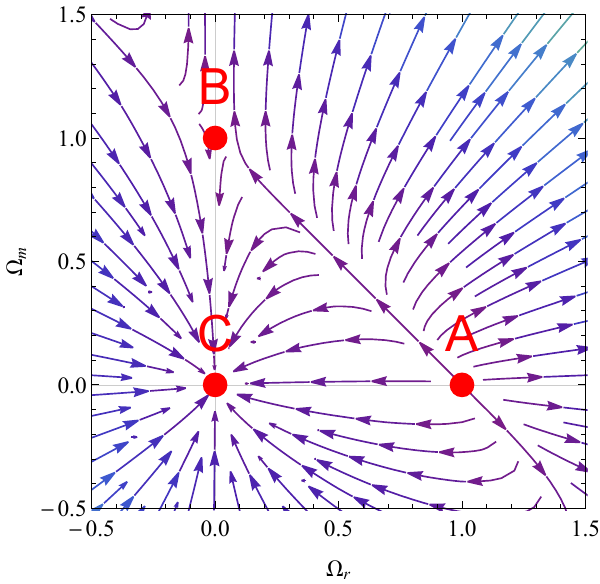} \quad
    \includegraphics[scale=0.53]{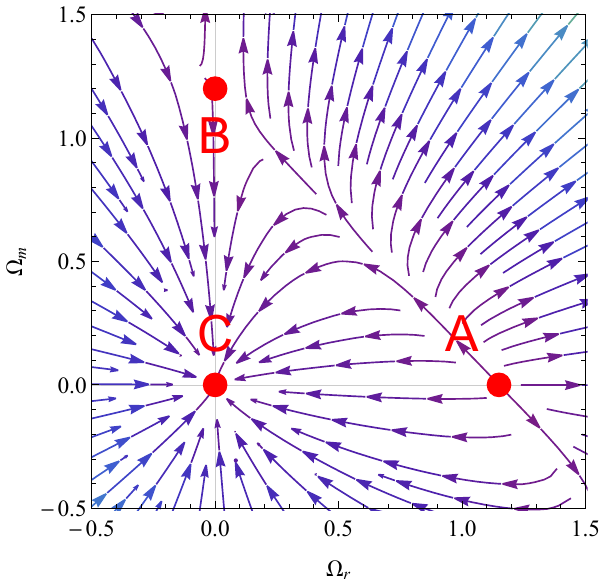} 
    \caption{Phase space trajectories of the dynamical system described by Eqs. \eqref{eq:simpR} to \eqref{eq:simpK} projected into the invariant submanifold $K=0$ for $X_0=-0.1$ (left panel), $X_0=0$ (middle panel), and $X_0=0.1$ (right panel). The middle panel corresponds to the GR limit. The critical points denoted by $\mathcal A$, $\mathcal B$, and $\mathcal C$ correspond respectively to radiation dominated, matter dominated, and exponentially accelerated solutions.}
    \label{fig:stream}
\end{figure*}

Equation \eqref{eq:dynfield} can also be used to compute the derivatives $\Phi'$ and $\Phi''$ and eliminate these derivatives from Eq. \eqref{eq:dynU}. The latter equation then becomes a constraint equation of the form
\begin{equation}
    \Psi\left(2Q-4-4K+2\Omega_r+3\Omega_m+6U\right)=0,
\end{equation}
which can be used to eliminate another dynamical variable from the system. Since the case $\Psi=0$ corresponds to the GR limit, we opt to consider $\Psi\neq 0$ and solve the constraint with respect to $Q$, which can then be used to eliminate this quantity from the system. The resultant dynamical system obtained under the assumptions outlined above takes the form
\begin{equation}\label{eq:simpR}
    \Omega'_r=\Omega_r\left(4\Omega_r+3\Omega_m-2K-6X_0-4\right),
\end{equation}
\begin{equation}\label{eq:simpM}
    \Omega'_m=\Omega_m\left(4\Omega_r+3\Omega_m-2K-6X_0-3\right).
\end{equation}
\begin{equation}\label{eq:simpK}
    K'=K\left(4\Omega_r+3\Omega_m-2K-6X_0-2\right),
\end{equation}
The system of Eqs. \eqref{eq:simpY} to \eqref{eq:simpM} feature a total of three invariant submanifolds, corresponding to the conditions $\Omega_r=0$, $\Omega_m=0$, and $K=0$ respectively. Since current observations seem to indicate that the geometry of the universe is approximately flat \cite{Planck:2018vyg}, i.e., $K\simeq0$, we consider a projection into that invariant submanifold and plot the trajectories in the phase space as a function of $\Omega_r$ and $\Omega_m$. In Fig. \ref{fig:stream}, we present the streamplots describing the phase space trajectories for different values of $X_0$. In these projections, one verifies that the radiation dominated solution, denoted by point $\mathcal A$ (this corresponds to the point $\mathcal A_1$ for $X_0\neq 0$ or to the convergence of points $\mathcal A_1$ and $\mathcal A_2$ if $X_0=0$), is a repeller in the phase space, the matter dominated solution, denoted by point $\mathcal B$ (this corresponds to the point $\mathcal B_1$ for $X_0\neq 0$ or to the convergence of points $\mathcal B_1$ and $\mathcal B_2$ if $X_0=0$), is a saddle point in the phase space, and the exponentially accelerated solution, denoted by point $\mathcal C$, is an attractor of the phase space. These properties remain independently of the value of $X_0$, with deviations from the GR case altering the phase space only quantitatively, while maintaining the overall qualitative behavior.

\subsection{Numerical integration}

The dynamical system described by Eqs. \eqref{eq:dynfield} to \eqref{eq:dynU} can also be numerically integrated under an appropriate set of initial conditions in order to extract the complete cosmological evolution of the model that satisfies the conditions imposed. Since we are interested in verifying if the hybrid metric-Palatini gravity theory allows for solutions comparable to those of the $\Lambda$CDM model, which are known solutions of standard GR, we start by imposing the GR limit in order to obtain the dynamical solutions for the density parameters $\Omega_r$, $\Omega_m$, and $U$ (the latter playing the role of the dark energy parameter $\Omega_\Lambda$ in GR), as well as the deceleration parameter $Q$. These parameters are subsequently used as reconstructive solutions to the general hybrid metric-Palatini model, in order to verify if the theory allows for those solutions while preserving the regularity and physical relevance of the scalar fields $\Phi$ and $\Psi$.

\begin{figure*}
    \centering
    \includegraphics[scale=0.85]{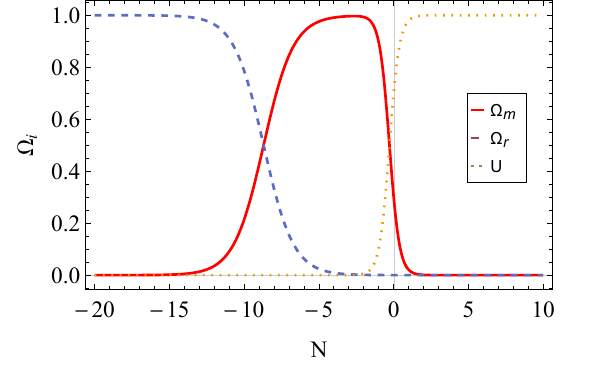}\quad 
    \includegraphics[scale=0.8]{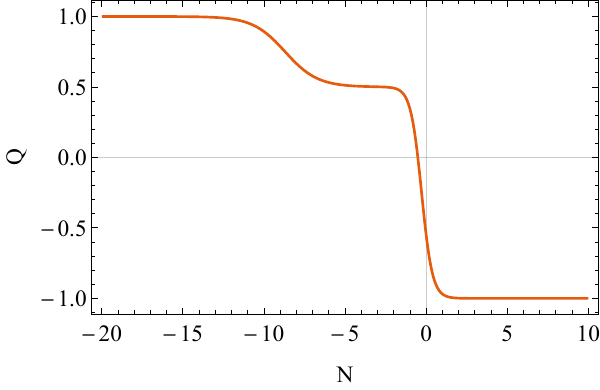}
    \caption{Density parameters $\Omega_r$, $\Omega_m$ and $U$ (left panel) and deceleration parameter $Q$ (right panel) obtained through a numerical integration of the dynamical system in Eqs. \eqref{eq:dynfield} to \eqref{eq:dynU} in the GR limit $\Phi-\Psi=1$ with $\Phi$ and $\Psi$ constant, and subjected to the initial conditions $K\left(0\right)=0$, $\Omega_r\left(0\right)=5 \times 10^{-5}$ and $\Omega_r\left(0\right)=0.3$.}
    \label{fig:parameters}
\end{figure*}

In order to numerically integrate the dynamical system, one must impose a set of initial conditions that are, ideally, consistent with experimental measurements. The current measurements of the cosmological parameters by the Planck satellite \cite{Planck:2018vyg} indicate that the geometry of the universe is approximately flat, i.e., $K\left(0\right)\simeq 0$, and that the density parameters for radiation and matter are approximately $\Omega_r\left(0\right)\simeq 5\times 10^{-5}$ and $\Omega_m\left(0\right)\simeq 0.3$, respectively. In the GR limit, i.e., $\Phi-\Psi=0$ with $\Phi=\Phi_0$ and $\Psi=\Psi_0$ constant values, the field equations given in Eqs. \eqref{eq:stfield1} and \eqref{eq:stfield2}, reduce to the forms
\begin{equation}\label{eq:const1}
    1+K=\Omega_r+\Omega_m+U,
\end{equation}
\begin{equation}\label{eq:const2}
    1+K-2Q=3U-\Omega_r.
\end{equation}
It follows from Eq. \eqref{eq:const1} that the current value of the density parameter $U$ is approximately $U\left(0\right)\simeq 0.69995$, whereas Eq. \eqref{eq:const2} can be used to eliminate the variable $Q$ from the dynamical system, which reduces to a system of three dynamical equations corresponding to Eqs. \eqref{eq:dynR} to \eqref{eq:dynU}. Performing a numerical integration of these three equations subjected to the initial conditions outlined above, one obtains the solutions for $\Omega_r$, $\Omega_m$, $U$, and $Q$ plotted in Fig. \ref{fig:parameters}. One observes that the universe starts from a radiation dominated era at early times, followed by a transition into a matter dominated era, and finally a transition to an exponentially accelerated phase, as expected. This model corresponds to the standard $\Lambda$CDM model.

Let us now verify if the solutions obtained in the GR limit, which are in agreement with the current cosmological measurements, are acceptable in the general hybrid metric-Palatini theory. Introducing the solutions plotted in Fig. \ref{fig:parameters} back into the dynamical system described by Eqs. \eqref{eq:dynfield} to \eqref{eq:dynU}, one verifies that Eqs. \eqref{eq:dynK} to \eqref{eq:dynM} are identically satisfied. This happens because these equations depend solely in the density parameters and deceleration parameter. On the other hand, Eqs. \eqref{eq:dynfield} and \eqref{eq:dynU} reduce to the following set of coupled ordinary differential equations for $\Phi$ and $\Psi$:
\begin{equation}\label{eq:dynlast1}
    \Phi+\Phi'=1+\Psi+\Psi'+\frac{\Psi'^2}{4\Psi},
\end{equation}
\begin{eqnarray}\label{eq:dynlast2}
    \Psi'\left[\left(4\Omega_r+3\Omega_m-6\right)\Psi'+\frac{\Psi'^2}{\Psi}-2\Psi''\right]=\nonumber \\
    =2\Psi\left(4\Omega_r+3\Omega_m-6\right)\left(\Phi'-\Psi'\right).
\end{eqnarray}
Equations \eqref{eq:dynlast1} and \eqref{eq:dynlast2} can be integrated numerically subjected to appropriate initial conditions in order to obtain the solutions for the scalar fields $\Phi$ and $\Psi$. Although there are no cosmological observations that could be used to constraint the initial conditions on $\Phi$ and $\Psi$, previous works have shown that a consistency with the solar system dynamics in the weak field regime of the hybrid metric-Palatini gravity can only be achieved if $\Phi\left(0\right)-\Psi\left(0\right)\simeq 1$ (see \cite{Rosa:2021lhc}). Furthermore, taking the early universe limit in which $\Omega_r\to1$ and $\Omega_m\to 0$, Eq. \eqref{eq:dynlast2} simplifies to
\begin{equation}\label{eq:dynlast3}
    2\Psi'\left(\Psi'+\Psi''\right)==\frac{\Psi'^3}{\Psi}.
\end{equation}
Equations \eqref{eq:dynlast1} and \eqref{eq:dynlast3} can be solved analytically to obtain the solutions
\begin{equation}
    \Phi\left(N\right)=e^{-N}\left(\Phi_0+e^N-1\right)+\frac{e^{-N}}{4\Psi_0}\left(e^N-1\right)\left(2\Psi_0+\Psi_1\right)^2,
\end{equation}
\begin{equation}
    \Phi\left(N\right)=\frac{e^{-2N}}{4\Psi_0}\left[\Psi_1-e^N\left(2\Psi_0+\Psi_1\right)\right]^2,
\end{equation}
where we have introduced the initial conditions $\Phi\left(0\right)=\Phi_0$, $\Psi\left(0\right)=\Psi_0$, and $\Psi'\left(0\right)=\Psi_1$. These solutions indicate that the scalar fields $\Phi$ and $\Psi$ are exponentially suppressed at early times during the radiation dominated phase. Furthermore, defining $\Xi=\Phi-\Psi$, one obtains
\begin{equation}
    \Xi\left(N\right)=\frac{e^{-2N}}{4\Psi_0}\left\{e^N\left[4\left(\Phi_0-\Psi_0+e^N-1\right)+\Psi_1^2\right]-\Psi_1^2\right\},
\end{equation}
which decays exponentially to $\Xi\to 1$. Thus, one expects not only that the present values of the fields $\Phi$ and $\Psi$ satisfy the requirement $\Phi-\Psi\simeq1$, which fulfills the constraints from weak field solar system dynamics, but also that the derivatives $\Phi'$ and $\Psi'$ should be small. 

\begin{figure*}
    \centering
    \includegraphics[scale=0.55]{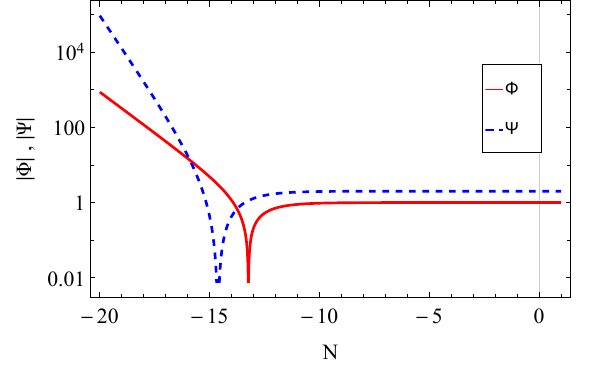}\quad
    \includegraphics[scale=0.55]{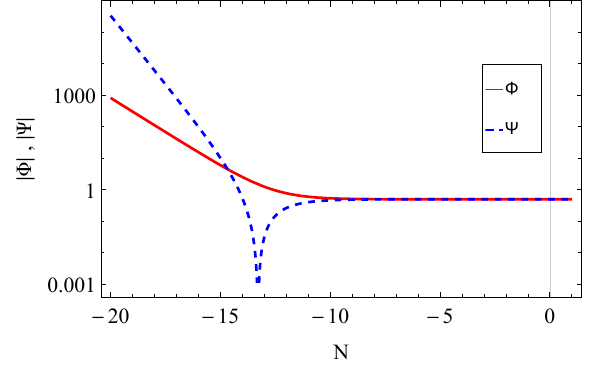}\quad
    \includegraphics[scale=0.55]{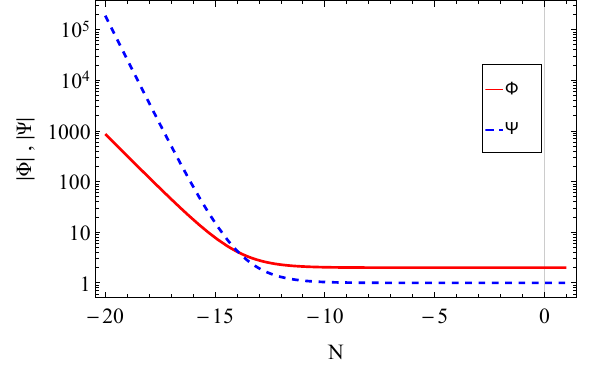}\\
    \includegraphics[scale=0.55]{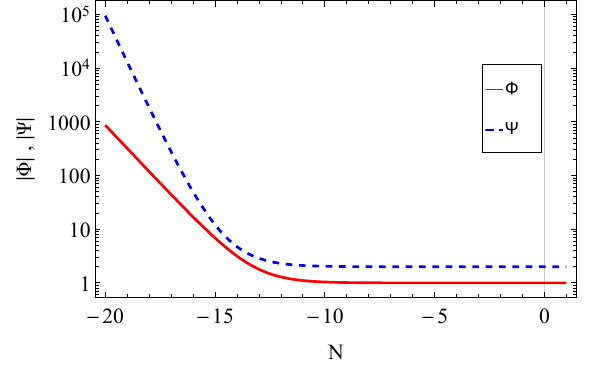}\quad
    \includegraphics[scale=0.55]{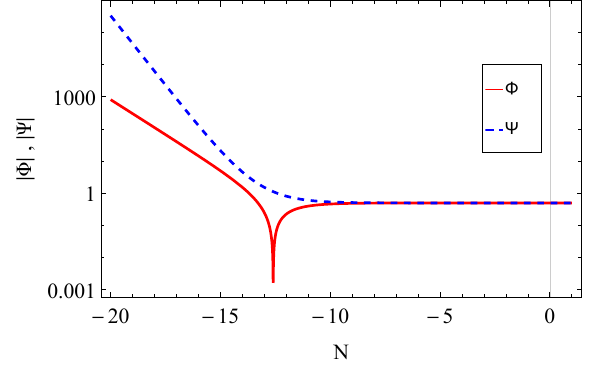}\quad
    \includegraphics[scale=0.55]{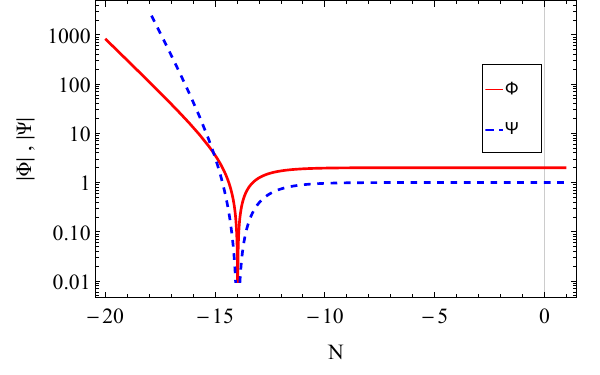}
    \caption{Solutions for $\Phi$ and $\Psi$ obtained through a numerical integration of Eqs. \eqref{eq:dynlast1} and \eqref{eq:dynlast2} subjected to the solutions for $\Omega_r$ and $\Omega_m$ presented in Fig. \ref{fig:parameters} and to the initial conditions $\{\Phi\left(0\right),\Psi\left(0\right)\}=\{-1,-2\}$ (left column), $\{\Phi\left(0\right),\Psi\left(0\right)\}=\{\frac{1}{2},-\frac{1}{2}\}$ (middle column), $\{\Phi\left(0\right),\Psi\left(0\right)\}=\{2,1\}$ (right column), $\Psi'\left(0\right)=-10^{-6}$ (top row), and $\Psi'\left(0\right)=10^{-6}$ (bottom row).}
    \label{fig:fields}
\end{figure*}

Following the considerations outlined above, consider the following four combinations of initial conditions: for the fields, we consider either that $\{\Phi\left(0\right),\Psi\left(0\right)\}=\{2,1\}$, $\{\Phi\left(0\right),\Psi\left(0\right)\}=\{-1,-2\}$, or $\{\Phi\left(0\right),\Psi\left(0\right)\}=\{\frac{1}{2},-\frac{1}{2}\}$, i.e., positive, negative, or mixed values of the fields, and we consider also $\Psi'\left(0\right)=\pm 10^{-6}$, i.e., positive or negative values of the derivatives. The absolute values of the solutions for $\Phi$ and $\Psi$ obtained through these numerical integrations are given in Fig. \ref{fig:fields}. It is noteworthy that the behavior of the fields is monotonic, and thus the fields are increasing whenever $\Psi' \left(0\right)>0$ and decreasing whenever $\Psi' \left(0\right)<0$. Consequently, the fields $\Phi$ and $\Psi$ change sign at some point in the past whenever $\Psi'\left(0\right)$ presents the same sign as $\Phi\left(0\right)$ or $\Psi\left(0\right)$, respectively. Summarizing, these results indicate that the scalar fields remain regular throughout the entire time evolution of the universe for a wide variety of parameter combinations compatible with the present weak field solar system dynamics. 

\section{Conclusions}\label{sec:concl}

In this work we used the formalism of dynamical systems to analyze the cosmological phase space of the scalar-tensor representation of a generalized form of hybrid metric-Palatini gravity and, more particularly, to reconstruct the $\Lambda$CDM model through a numerical integration of the resultant dynamical system. The use of the scalar-tensor representation of the theory greatly reduces the complexity of the analysis in comparison with previous works where the geometrical representation was considered. Indeed, in this representation one can recur to the usual density parameters for the $\Lambda$CDM model used in GR, which allows for a direct comparison of the model with observational data. 

This work improves on the existent literature on the topic (see \cite{Carloni:2015bua,Tamanini:2013ltp,Rosa:2019ejh}) in several fronts. First, we have explicitly considered the matter distribution to be a combination of two relativistic fluids describing radiation and dust, a trait that allows one to obtain cosmological solutions featuring the necessary radiation and matter dominated epochs to explain phenomena e.g. nucleosynthesis and large scale structure formation, in contrast with previous works where the equation of state for the matter components was considered arbitrary. Second, the solutions obtained in this work follow initial conditions that are in agreement with both the cosmographic measurements of the Planck satellite and the weak-field solar system dynamics, thus being of a larger physical relevance. Third, the analysis carried in this work is model independent, i.e., the dynamical system was written in a convenient form that allows for the solutions to be obtained without the need to specify \textit{a priori} an explicit form of the potential $V\left(\varphi,\psi\right)$ (or, equivalently an explicit form of the function $f\left(R,\mathcal R\right)$. 

Our analysis indicates that the hybrid metric-Palatini theory of gravity allows for cosmological solutions qualitatively identical to those of the $\Lambda$CDM model in GR, starting from a radiation dominated epoch at early times, transitioning to a matter dominated epoch, and finally undergoing a transition into an exponentially accelerated phase. Indeed, deviations from GR in this framework alter only quantitatively the phase space, while preserving the overall qualitative behavior. The interaction potential of the scalar fields was shown to effectively play the role of dark energy and being the dominant quantity during the exponentially accelerated period. Furthermore, our analysis indicates that the deviations of the theory from GR are exponentially suppressed during the radiation dominated phase, with the scalar fields rapidly approaching values consistent with those necessary to fulfill the solar system constraints. These models can thus be interpreted as equivalent to a dynamical dark energy model that asymptotically approaches the behavior of a cosmological constant at late times.

Summarizing, our results indicate that the hybrid metric-Palatini gravity can effectively explain the late-time cosmic acceleration without the necessity of an explicit dark energy component. However, and similarly to what happens with a wide variety of similar modified theories of gravity, it is noteworthy that, at the level analyzed in this work, the models presented are virtually indistinguishable from the $\Lambda$CDM model, which could potentially render them unfalsifiable. Further analysis on these models and the development of some observable that allows one to resolve this degeneracy is justified, although out of the scope of this work.

\begin{acknowledgments}
J.L.R. acknowledges the European Regional Development Fund and the programme Mobilitas Pluss for financial support through Project No.~MOBJD647, project No.~2021/43/P/ST2/02141 co-funded by the Polish National Science Centre and the European Union Framework Programme for Research and Innovation Horizon 2020 under the Marie Sklodowska-Curie grant agreement No. 94533, Fundação para a Ciência e Tecnologia through project number PTDC/FIS-AST/7002/2020, and Ministerio de Ciencia, Innovación y Universidades (Spain), through grant No. PID2022-138607NB-I00.
\end{acknowledgments}


\end{document}